\documentclass[aps,prl,preprint,superscriptaddress]{revtex4}

\usepackage{graphicx}
\begin{document}

\title{Imaging nonequilibrium atomic vibrations with x-ray diffuse scattering}

\author{M. Trigo, Y. M. Sheu, J. Chen, V. H. Vishwanath, T. Graber, R. Henning and D. A. Reis}
\affiliation{}


\date{\today}

\begin{abstract}
\end{abstract}


\maketitle

For over a century, x-ray scattering has been the most powerful tool for determining the equilibrium structure of crystalline materials. Deviations from perfect periodicity, for example due to thermal motion of the atoms, reduces the intensity of the Bragg peaks as well as produces structure in the diffuse scattering background~\cite{james_book,warren_book}.  Analysis of the thermal diffuse scattering (TDS) had been used to determine interatomic force constants and phonon dispersion in relatively simple cases~\cite{joynson1954,jacobsen1955,cole1953,cole1952} before inelastic neutron scattering became the preferred technique to study lattice dynamics~\cite{brockhouse1955}.  With the advent of intense synchrotron x-ray sources, there was a renewed interest in TDS for measuring phonon dispersion~\cite{holt1999}. The relatively short x-ray pulses emanating from these sources also enables the measurement of phonon dynamics in the time domain. Prior experiments on nonequilibrium phonons were either limited by time-resolution \cite{mcwhan1982,chapman1984} and/or to relatively long wavelength excitations~\cite{leroux1975,reis2007,trigo_reisMRS2010}.   Here we present the first images of nonequilibrium phonons throughout the Brillouin zone in photoexcited III-V semiconductors, indium-phosphide (InP) and indium-antimonide (InSb), using picosecond time-resolved diffuse scattering. In each case, we find that the lattice remain out of equilibrium for several hundred picoseconds up to nanoseconds after laser excitation. The non-equilibrium population is dominated by transverse acoustic phonons which in InP are directed along high-symmetry directions.
The results have wide implications for the detailed study of electron-phonon and phonon-phonon coupling in solids.

Experiments were performed at the BioCARS beamline of the Advanced Photon Source at Argonne National Laboratory. A dual undulator setup provided an x-ray photon flux at the sample of $\sim 10^{10}$~photons/pulse with photon energy of $12 - 15$~keV in a $ < 4~\%$ bandwidth.
Single $100$~ps x-ray pulses were isolated by a high-speed chopper at a rate of $40$~Hz. 
A Ti:sapphire regenerative amplifier synchronized to the accelerator RF pumped an optical parametric amplifier to produce visible pulses $1.2$~ps in duration. The pulse energy at the sample was $\sim 62~\mu$J focused on a $0.14 \times 2.1$~mm spot.
The samples were single crystals of InP and InSb with the $(0 0 1)$ direction perpendicular to the wafer surface. The crystals were oriented at grazing incidence so that the x-rays were incident primarily along the $(1 0 0)$ direction. Typical incidence angles of $\alpha = 0.3^{\rm o}$ were used to match the laser excitation depth with the volume probed by the x-rays. 
A large area detector (MarCCD 165~mm diameter) placed $64$~mm behind the sample collects the x-rays with scattering angles of up to $2 \theta \sim 51^{\rm o}$. 
In choosing the x-ray energy and crystal orientation, special care was taken to avoid Bragg reflections on the detector. 

\begin{figure}
\includegraphics[scale=0.6]{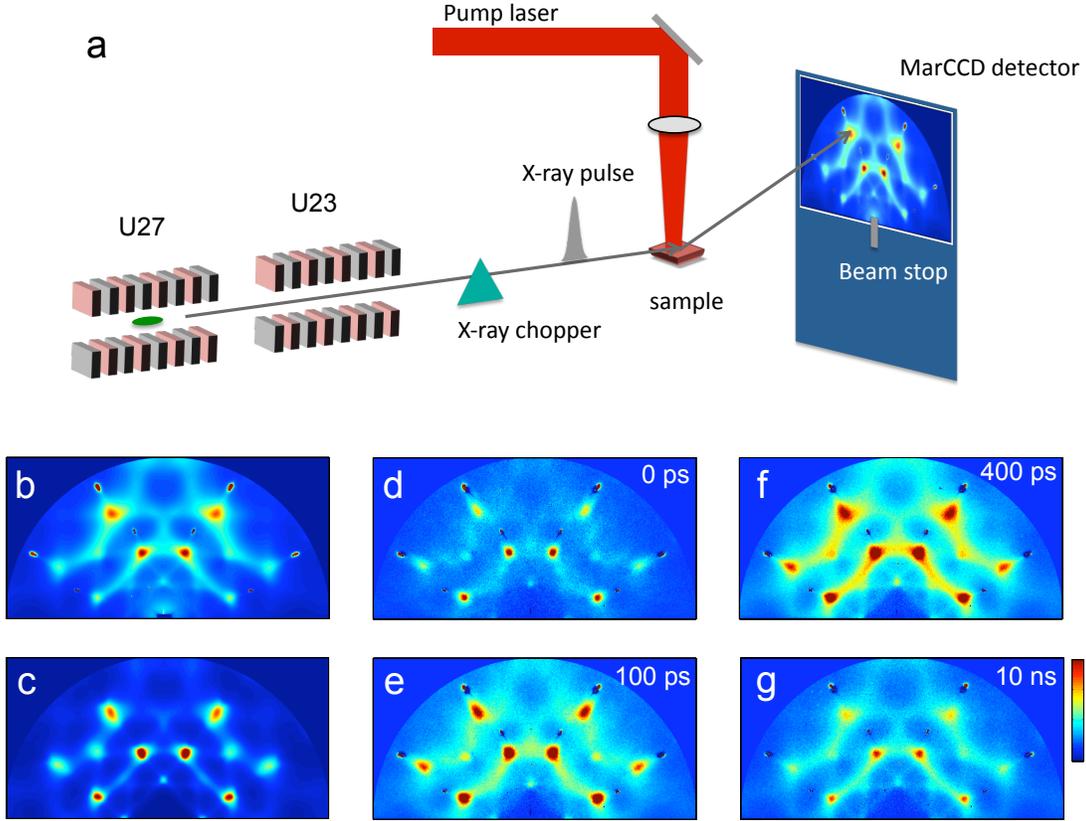}%
\caption{(a) Schematics of the experimental setup for time-resolved diffuse scattering at BioCARS beamline at APS. A single X-ray pulse produced by two collinear undulators is isolated by a high speed chopper. A CCD detector 165 mm in diameter placed 64 mm behind the sample collects the scattered photons. A $\sim 1$~ps laser pulse from a tunable optical parametric amplifier is synchronized with the x-ray pulse. (b) room temperature TDS image of InP at $15$~keV oriented with $(1 0 0)$ parallel to the x-ray direction.  (c) Calculated TDS from a Born-von-K\'arm\'an model with six nearest-neighbors forces at $15$~keV using a $2~\%$ Gaussian energy spread. (d) - (g) differences between laser-exposed and laser-off frames at delays $\Delta t = 0$, $100$, $400$~ps, and $10$~ns, respectively.
 \label{Fig1}}
 \end{figure}

Figure \ref{Fig1} (b) shows a static TDS pattern from InP at room temperature taken with $15$~keV x-ray photons. 
The upper half of the image is shown, the lower half being blocked by the sample.
The large area detector captures a solid angle of the Ewald sphere in reciprocal space projected onto a plane. Each pixel has associated a different scattering vector ${\bf Q}$ that depends on the x-ray energy and the sample-detector distance.
The intensity pattern originates primarily from (inelastic) scattering from thermally activated phonons and shows structure that reflects the details of the phonon dispersion in the material and the phonon population at a given temperature~\cite{chiang2005}.

Ultrafast laser excitation of the material initially produces a non-equilibrium state where phonons may not be populated according to the Bose factor. This transient state eventually leads to a new equilibrium where the lattice is left at a higher temperature.
These  perturbations of the phonon population modify the equilibrium diffuse scattering pattern shown in Fig. 1 (b). As a function of laser-x-ray delay, $t_i$, we record a frame with laser exposure, $I(t_j)$, and a dark frame, $I({\rm off})$.
Figures~\ref{Fig1} $(d) - (g)$  show the change in the image upon laser excitation for several representative time delays [i.e. $I(t_j) - I({\rm off})$]. Each data frame $I(t_j)$ is obtained by averaging 100 individual images with 100 single-pulse exposures per image. 
Initially, the areas near the zone-centers brighten at $t=0$~ps (d) while the scattering keeps increasing several hundred ps after excitation (e) and (f), before decreasing  after a few ns (g), eventually returning to equilibrium. 
The laser and x-rays are coincident at $t = 0$~ps (d), however the initial increase in scattering takes $\sim 100$~ps to develop (e). 

We consider the intensity of one-phonon thermal diffuse scattering given by\cite{chiang2005}
\begin{equation}\label{eqtds}
I_{1}({\bf Q}) \propto \sum_j \frac{1}{\omega_j({\bf q})} \left( n_{j}({\bf q})+\frac{1}{2}\right) \left| F_j({\bf Q})  \right|^2,
\end{equation}
where $\omega_j({\bf q})$ is the frequency of the phonon mode in branch $j$ with reduced wavevector ${\bf q}$, $n_{j}({\bf q})$ is the phonon population and $F_j({\bf Q})$ is given by
\begin{equation}\label{eqfq}
F_j({\bf Q}) = \sum_s \frac{f_s}{\sqrt{m_s}} e^{-M_s} ({\bf Q} \cdot \hat{\bf e}_{s,j,{\bf q}}) e^{-i {\bf K}_Q\cdot {\bf r}_s}.
\end{equation}
In this expression, $f_s$, $m_s$, $M_s$ are the atomic scattering factor, the mass, and the Debye-Waller factor~\cite{vetelino1972} of atom $s$ at position ${\bf r}_s$, $\hat{\bf e}_{s,j,{\bf q}}$ is the phonon polarization vector, and ${\bf K}_Q$ is the closest reciprocal lattice vector to ${\bf Q}$, i.e. ${\bf Q} = {\bf q} + {\bf K}_Q$. 
In Eq.~(\ref{eqtds}) we omitted additional slowly varying factors from the polarization dependence of the scattering as well as geometrical factors~\cite{chiang2005}.

In thermal equilibrium, $n_{j}({\bf q}) = \frac{1}{e^{-\omega_j({\bf q})/k_B T}-1}$, which, for the low frequency limit gives $n_{j}({\bf q}) \propto k_B T/\omega_j({\bf q})$. This multiplied by the
$1/{\omega_j({\bf q})}$ factor in Eq. (\ref{eqtds})  gives a strong contribution from acoustic modes near the center of the Brillouin zone where their frequency goes to zero. These low-frequency phonons give rise to the four brightest diffuse spots in Fig. \ref{Fig1} (b). Directions in reciprocal space where the dispersion relation is soft appear in Fig.~\ref{Fig1} (b) as bright lines connecting the near-zone-center spots.
The quantity $F_j({\bf Q})$ in Eq. (\ref{eqfq}) has the form of a structure factor modified by the dot product $({\bf Q} \cdot \hat{\bf e}_{s,j,{\bf q}})$, which selects phonon modes polarized along the scattering vector.
A slight deviation from the four-fold symmetry of the crystal is expected between the vertical and horizontal directions due to the horizontally polarized x-rays.
The eight small spots of a few pixels in diameter in Fig. \ref{Fig1} (b) are crystal surface truncation rods which appear due to the grazing incidence geometry and are well known from surface diffraction~\cite{robinson1986}.  

Equations (\ref{eqtds}) and (\ref{eqfq}) can be used to compute the TDS if the phonon frequencies {\it and} eigenvectors are known. 
{The inverse problem, i.e. obtaining  $\omega_j({\bf q})$ and $\hat{\bf e}_{s,j,{\bf q}}$ from $I({\bf Q})$ can be done model-free only in simple cases~\cite{bosak2008}}. 
To gain insight on the different phonon contributions we implemented a Born-von K\'arm\'an model of the lattice dynamics with force constants up to six nearest neighbors\cite{soma1983,kagaya1984}. InP and InSb crystallize in the zincblende structure with two atoms per unit cell. Six phonon branches and eigen-displacements, three acoustic and three optical, are obtained.  
This model ignores the ionic component of the forces and thus does not reproduce the phonon dispersion with all detail, particularly it does not give a splitting of the LO and TO phonons at zone-center. 
However the insight gained is extremely valuable for understanding the static diffuse images particularly in separating the contributions from transverse (TA) and longitudinal (LA) acoustic phonons.
Figure \ref{Fig1} (c) shows a calculated TDS intensity pattern using this force constant model.
This calculation agrees well with the static measurements and reproduces the main features. More importantly, as we show later, it proves useful in understanding the time-resolved data.

To separate the different time-scales we performed singular value decomposition (SVD) of the time-resolved data. 
This method has been applied successfully to a wide set of problems from small angle scattering~\cite{okamoto2003} and the analysis of spectroscopic data~\cite{henry1997} to time-resolved Laue diffraction of protein crystals~\cite{schmidt2002}.
The procedure is as follows. We write each difference frame $I(t_j) - I({\rm off})$ as a column vector ${\bf x}_j$ and construct the matrix ${\bf X}$ whose columns are the vectors ${\bf x}_j$. Thus, adjacent columns of ${\bf X}$ correspond to consecutive frames in the time-delay series. The SVD states that ${\bf X} = {\bf U}{\bf S}{\bf V}^T$, where ${\bf S} = {\rm diag}(s_i)$ is the diagonal matrix of singular values with $s_i \geq 0$, and ${\bf U}$ and ${\bf V}$ are orthogonal matrices whose columns,  ${\bf u}_j$ and  ${\bf v}_j$, are the left and right singular vectors, respectively.
The vectors ${\bf u}_j$ represent time-independent ``populations'', and form a complete orthonormal basis. The vectors ${\bf v}_j$ contain the time-dependent information of each of the populations and also form an orthonormal basis. The singular values $\{s_i\}$ act as weight factors for these vectors and represent the amount of ``signal'' contained in each singular vector in the reconstructed data.

\begin{figure}
\includegraphics[scale=0.6]{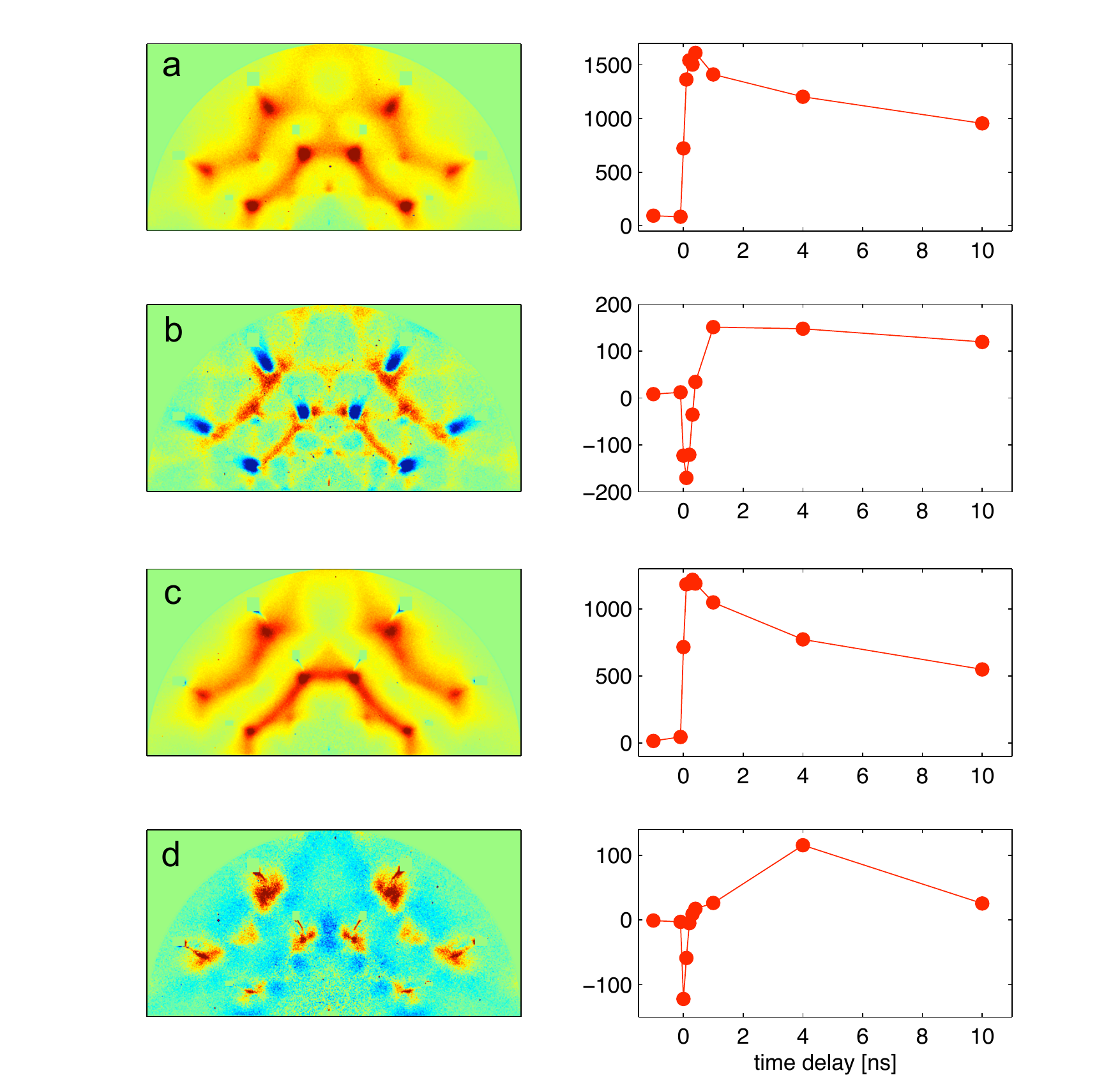}%
\caption{Results of the singular value decomposition analysis of the time-resolved data. The procedure decomposes the data as a sum of individual images (left) each with a corresponding time dependence (right). (a) and (b) correspond to the two most significant singular values for InP and (c) and (d) for InSb.
\label{Fig2}}
\end{figure}

In Fig. \ref{Fig2} we summarize the results of the SVD analysis for both InP [(a) and (b)] and InSb [(c) and (d)]. The truncation rods were masked out to avoid artifacts from saturated pixels. Due to the larger lattice constant, the data for InSb were taken at $13$~keV to match the same reciprocal space covererage on the detector as InP at $15$~keV. On the left we show the (time-independent) left singular vectors (lSV) for the two most significant singular values (SV) with the corresponding  (time-dependent) right singular vectors (rSV) scaled by the corresponding SV on the right. 
For both materials, much of the signal is contained in the first SV [Fig. \ref{Fig2} (a) and (c)], the time dependence of which consists of a sharp increase in the diffuse scattering that decays in few ns time scale (Fig. \ref{Fig2} (a) and (c) right panels).
It is worth noting here the clear differences between InP and InSb, particularly the lack of scattering intensity in InSb near the top center of the first lSV in Fig. \ref{Fig2} (c), which is also present in the dark frame (not shown). This is due to the small structure factor for the (204) reflection in InSb, which is not small in InP due to the large difference in the atomic scattering factors.  
The shape of the image from this lSV resembles the equilibrium image, which suggests this lSV corresponds to  lattice heating and diffusion that decays within a time-scale of few ns. 

The second singular vectors however show a more complex behavior both in reciprocal space as well as in the time domain. In the case of InP, the lSV [Fig. \ref{Fig2} (b) (left)] shows sharp bright and dark areas indicating that parts of the image brighten while other areas decrease their scattering. The fact that the rSV [Fig. \ref{Fig2} (b) (right)] peaks at $t = 400$~ps means that the image on the left continues increasing after the initial excitation which indicates that non-equilibrium transient dynamics lead to a delayed heating and cooling of different parts of the Brillouin-zone. The same delayed behavior is observed in InSb as seen in Fig. \ref{Fig2} (d) (right), which shows a peak at $\sim 4$~ns. We attribute this behavior to relaxation of non-equilibrium phonon populations. However, note that from the way the population $n_{j}({\bf q})$ and frequency $\omega_{j}({\bf q})$ enter in Eq. (\ref{eqtds}), it is difficult to distinguish a population increase from a softening of the phonon frequencies.

\begin{figure}
\includegraphics[scale=0.6]{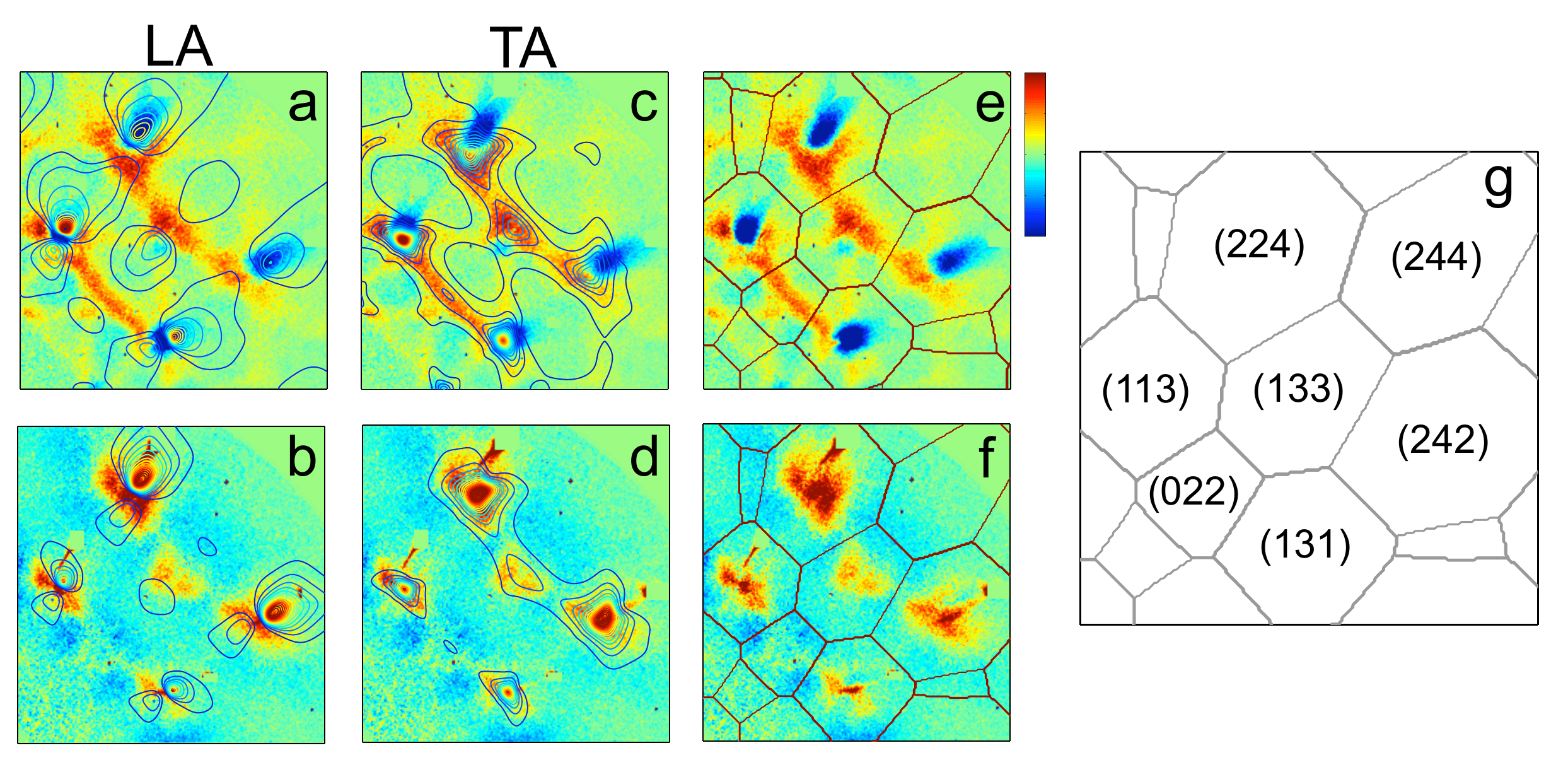}%
\caption{Contribution to the TDS from TA and LA branches calculated using Eq.~(\ref{eqtds}). The images correspond to an expanded view of the second eigenvectors in Fig.~\ref{Fig2} (b) (top row) and Fig.~\ref{Fig2} (d) (bottom row). The contours in (a) and (b) show the calculated contribution from LA phonons in InP and InSb, respectively. (c) and (d) show the contribution from TA phonons for InP and InSb, respectively. (e) and (f) shows the same images superimposed with the Brillouin-zone boundaries for InP and InSb. (g) Miller indices for selected Brillouin-zones of InSb.
\label{Fig3}}
\end{figure}

Although the (equilibrium) TDS distributions for InP and InSb are similar, as well as  the first singular vector in Fig.~\ref{Fig2} (a) and (c), which reflects similarities in the phonon dispersion, the difference in the second lSV is quite remarkable and cannot be explained by differences in the equilibrium phonon dispersion alone.
It is instructive to separate the contribution from different phonon branches to the diffuse scattering. Each phonon branch contributes one term to Eq.~ (\ref{eqtds}), which we compute using the Born-von K\'arm\'an model described above.
The top and bottom rows of Fig.~\ref{Fig3} show a zoomed view of the second lSV in Fig.~\ref{Fig2} (b) and (d), respectively. Superimposed, we show contour plots that represent the calculated contribution to Eq.~(\ref{eqtds}) from the LA [(a) and (b)] and the TA  [(c) and (d)] branches separately. 
As stated earlier, the largest contribution to the scattering comes from areas that are close to the center of the Brillouin-zone, where the contours lines are denser. This is further confirmed by superimposing the images with the corresponding Brillouin-zone boundaries as shown in (e) and (f).

From Fig.~\ref{Fig3} we see that in InP the areas with strong contribution from TA phonons [Fig.~\ref{Fig3} (c)] appear bright in the lSV, while areas with more contribution from the LA branch [Fig.~\ref{Fig3} (a)] are dark. This lSV has the time-dependence shown in the right panel of Fig.~\ref{Fig2} (b), which means that even at time-delays $t > 400$~ps TA phonons are still populating while the LA branch decreases.
In InSb the situation is different. The LA branch shows no appreciable decrease in scattering above the noise level [Fig.~\ref{Fig3} (b)], while the TA branches populate delayed [Fig.~\ref{Fig3} (d)].
To further identify which phonon modes are involved within the Brillouin-zone, we sketch in Fig.~\ref{Fig3} (g) the Miller indices of selected  Brillouin-zones for InSb at $13$~keV. Note that at this energy the number and size of Brillouin zones covered by the detector in InSb is similar to InP at $15$~keV. 
Using the diagram in Fig.~\ref{Fig3} (g) we can assign the InP TA phonons in Fig.~\ref{Fig3} (c) to modes along high-symmetry $(111)$ and $(010)$ or equivalent directions. 
InSb on the other hand shows a less directional scattering from the TA branches, indicating that the population is distributed more isotropically within the Brillouin-zone. We attribute this difference to the large gap between the optical and acoustic modes in InP, which leads to less available phonon decay channels~\cite{debernardi1998}.

In conclusion, we have shown that time-resolved non-thermal x-ray diffuse scattering yields heretofore inaccessible information of the non-equilibrium dynamics of solids under ultrafast excitation.  
InP and InSb show complex non-equilibrium redistributions of the scattered intensity throughout the Brillouin-zone that are clearly separated by singular value decomposition. This non-equilibrium lattice persists up to several hundred ps. 
This behavior is interpreted using a Born-von K\'arm\'an model which reveals that TA phonons populate several hundred ps after initial laser excitation and, in the case of InP, is very directional and appears concomitant with a decrease in the scattering by LA phonons. In the current study, we are limited to dynamics that occur on more than 100 ps time-scale due to the x-ray pulse duration.
We note that measurements of the initial emission of hot phonons by the photoexcited carriers and their subsequent anharmonic decay will become possible with new x-ray sources such as the Linac Coherent Light Source at SLAC.  As these new sources of ultra-bright femtosecond x-ray pulses come online, x-ray diffuse scattering becomes a promissing approach to explore the ultrafast dynamics of solids under laser excitation at the femtosecond time-scale. 

We acknowledge S. Ghimire for experimental assistance and H. Dosch and T. C. Chiang for fruitful discussions.
This research is supported through the PULSE Institute at the SLAC National Accelerator Laboratory by the U.S. Department of Energy, Office of Basic Energy Sciences.

\begin{thebibliography}{23}
\expandafter\ifx\csname natexlab\endcsname\relax\def\natexlab#1{#1}\fi
\expandafter\ifx\csname bibnamefont\endcsname\relax
  \def\bibnamefont#1{#1}\fi
\expandafter\ifx\csname bibfnamefont\endcsname\relax
  \def\bibfnamefont#1{#1}\fi
\expandafter\ifx\csname citenamefont\endcsname\relax
  \def\citenamefont#1{#1}\fi
\expandafter\ifx\csname url\endcsname\relax
  \def\url#1{\texttt{#1}}\fi
\expandafter\ifx\csname urlprefix\endcsname\relax\def\urlprefix{URL }\fi
\providecommand{\bibinfo}[2]{#2}
\providecommand{\eprint}[2][]{\url{#2}}

\bibitem[{\citenamefont{James}(1954)}]{james_book}
\bibinfo{author}{\bibfnamefont{R.}~\bibnamefont{James}},
  \emph{\bibinfo{title}{Optical Principles of the Diffraction of X-Rays}}
  (\bibinfo{publisher}{G. Bell}, \bibinfo{address}{London},
  \bibinfo{year}{1954}).

\bibitem[{\citenamefont{Warren}(1969)}]{warren_book}
\bibinfo{author}{\bibfnamefont{B.~E.} \bibnamefont{Warren}},
  \emph{\bibinfo{title}{X-Ray Diffraction}} (\bibinfo{publisher}{Dover},
  \bibinfo{address}{New York}, \bibinfo{year}{1969}).

\bibitem[{\citenamefont{Joynson}(1954)}]{joynson1954}
\bibinfo{author}{\bibfnamefont{R.~E.} \bibnamefont{Joynson}},
  \bibinfo{journal}{Phys. Rev.} \textbf{\bibinfo{volume}{94}},
  \bibinfo{pages}{851} (\bibinfo{year}{1954}).

\bibitem[{\citenamefont{Jacobsen}(1955)}]{jacobsen1955}
\bibinfo{author}{\bibfnamefont{E.~H.} \bibnamefont{Jacobsen}},
  \bibinfo{journal}{Phys. Rev.} \textbf{\bibinfo{volume}{97}},
  \bibinfo{pages}{654} (\bibinfo{year}{1955}).

\bibitem[{\citenamefont{Cole}(1953)}]{cole1953}
\bibinfo{author}{\bibfnamefont{H.}~\bibnamefont{Cole}},
  \bibinfo{journal}{Journal of Applied Physics} \textbf{\bibinfo{volume}{24}},
  \bibinfo{pages}{482} (\bibinfo{year}{1953}).

\bibitem[{\citenamefont{Cole and Warren}(1952)}]{cole1952}
\bibinfo{author}{\bibfnamefont{H.}~\bibnamefont{Cole}} \bibnamefont{and}
  \bibinfo{author}{\bibfnamefont{B.~E.} \bibnamefont{Warren}},
  \bibinfo{journal}{Journal of Applied Physics} \textbf{\bibinfo{volume}{23}},
  \bibinfo{pages}{335} (\bibinfo{year}{1952}).

\bibitem[{\citenamefont{Brockhouse and Stewart}(1955)}]{brockhouse1955}
\bibinfo{author}{\bibfnamefont{B.~N.} \bibnamefont{Brockhouse}}
  \bibnamefont{and} \bibinfo{author}{\bibfnamefont{A.~T.}
  \bibnamefont{Stewart}}, \bibinfo{journal}{Phys. Rev.}
  \textbf{\bibinfo{volume}{100}}, \bibinfo{pages}{756} (\bibinfo{year}{1955}).

\bibitem[{\citenamefont{{M. Holt} et~al.}(1999)\citenamefont{{M. Holt}, {Z.
  Wu}, {Hawoong Hong}, {P. Zschack}, {P. Jemian}, { J. Tischler}, {Haydn Chen},
  and { T.-C. Chiang}}}]{holt1999}
\bibinfo{author}{\bibnamefont{{M. Holt}}}, \bibinfo{author}{\bibnamefont{{Z.
  Wu}}}, \bibinfo{author}{\bibnamefont{{Hawoong Hong}}},
  \bibinfo{author}{\bibnamefont{{P. Zschack}}},
  \bibinfo{author}{\bibnamefont{{P. Jemian}}}, \bibinfo{author}{\bibnamefont{{
  J. Tischler}}}, \bibinfo{author}{\bibnamefont{{Haydn Chen}}},
  \bibnamefont{and} \bibinfo{author}{\bibnamefont{{ T.-C. Chiang}}},
  \bibinfo{journal}{Phys. Rev. Lett.} \textbf{\bibinfo{volume}{83}},
  \bibinfo{pages}{3317} (\bibinfo{year}{1999}).

\bibitem[{\citenamefont{McWhan et~al.}(1982)\citenamefont{McWhan, Hu, Chin, and
  Narayanamurti}}]{mcwhan1982}
\bibinfo{author}{\bibfnamefont{D.~B.} \bibnamefont{McWhan}},
  \bibinfo{author}{\bibfnamefont{P.}~\bibnamefont{Hu}},
  \bibinfo{author}{\bibfnamefont{M.~A.} \bibnamefont{Chin}}, \bibnamefont{and}
  \bibinfo{author}{\bibfnamefont{V.}~\bibnamefont{Narayanamurti}},
  \bibinfo{journal}{Phys. Rev. B} \textbf{\bibinfo{volume}{26}},
  \bibinfo{pages}{4774} (\bibinfo{year}{1982}).

\bibitem[{\citenamefont{Chapman et~al.}(1984)\citenamefont{Chapman, Hsieh, and
  Colella}}]{chapman1984}
\bibinfo{author}{\bibfnamefont{L.~D.} \bibnamefont{Chapman}},
  \bibinfo{author}{\bibfnamefont{S.~M.} \bibnamefont{Hsieh}}, \bibnamefont{and}
  \bibinfo{author}{\bibfnamefont{R.}~\bibnamefont{Colella}},
  \bibinfo{journal}{Phys. Rev. B} \textbf{\bibinfo{volume}{30}},
  \bibinfo{pages}{1094} (\bibinfo{year}{1984}).

\bibitem[{\citenamefont{LeRoux et~al.}(1975)\citenamefont{LeRoux, Colella, and
  Bray}}]{leroux1975}
\bibinfo{author}{\bibfnamefont{S.~D.} \bibnamefont{LeRoux}},
  \bibinfo{author}{\bibfnamefont{R.}~\bibnamefont{Colella}}, \bibnamefont{and}
  \bibinfo{author}{\bibfnamefont{R.}~\bibnamefont{Bray}},
  \bibinfo{journal}{Phys. Rev. Lett.} \textbf{\bibinfo{volume}{35}},
  \bibinfo{pages}{230} (\bibinfo{year}{1975}).

\bibitem[{\citenamefont{Reis and Lindenberg}(2007)}]{reis2007}
\bibinfo{author}{\bibfnamefont{D.~A.} \bibnamefont{Reis}} \bibnamefont{and}
  \bibinfo{author}{\bibfnamefont{A.~M.} \bibnamefont{Lindenberg}}, in
  \emph{\bibinfo{booktitle}{Light Scattering in Solids IX, Topics in Applied
  Physics}}, edited by
  \bibinfo{editor}{\bibfnamefont{M.}~\bibnamefont{Cardona}} \bibnamefont{and}
  \bibinfo{editor}{\bibfnamefont{R.}~\bibnamefont{Merlin}}
  (\bibinfo{publisher}{Springer}, \bibinfo{year}{2007}), vol.
  \bibinfo{volume}{108}, pp. \bibinfo{pages}{371--422}.

\bibitem[{\citenamefont{Trigo and Reis}(2010)}]{trigo_reisMRS2010}
\bibinfo{author}{\bibfnamefont{M.}~\bibnamefont{Trigo}} \bibnamefont{and}
  \bibinfo{author}{\bibfnamefont{D.}~\bibnamefont{Reis}}, \bibinfo{journal}{MRS
  Bulletin} \textbf{\bibinfo{volume}{35}} (\bibinfo{year}{2010}).

\bibitem[{\citenamefont{{Ruqing Xu} and {Tai C. Chiang}}(2005)}]{chiang2005}
\bibinfo{author}{\bibnamefont{{Ruqing Xu}}} \bibnamefont{and}
  \bibinfo{author}{\bibnamefont{{Tai C. Chiang}}}, \bibinfo{journal}{Z.
  Kristallogr.} \textbf{\bibinfo{volume}{220}}, \bibinfo{pages}{1009}
  (\bibinfo{year}{2005}).

\bibitem[{\citenamefont{{J. F. Vetelino} et~al.}(1972)\citenamefont{{J. F.
  Vetelino}, {S. P. Gaur}, and {S. S. Mitra}}}]{vetelino1972}
\bibinfo{author}{\bibnamefont{{J. F. Vetelino}}},
  \bibinfo{author}{\bibnamefont{{S. P. Gaur}}}, \bibnamefont{and}
  \bibinfo{author}{\bibnamefont{{S. S. Mitra}}}, \bibinfo{journal}{Phys. Rev.
  B} \textbf{\bibinfo{volume}{5}}, \bibinfo{pages}{2360}
  (\bibinfo{year}{1972}).

\bibitem[{\citenamefont{Robinson}(1986)}]{robinson1986}
\bibinfo{author}{\bibfnamefont{I.~K.} \bibnamefont{Robinson}},
  \bibinfo{journal}{Phys. Rev. B} \textbf{\bibinfo{volume}{33}},
  \bibinfo{pages}{3830} (\bibinfo{year}{1986}).

\bibitem[{\citenamefont{{Alexei Bosak} and {Dmitry
  Chernyshov}}(2008)}]{bosak2008}
\bibinfo{author}{\bibnamefont{{Alexei Bosak}}} \bibnamefont{and}
  \bibinfo{author}{\bibnamefont{{Dmitry Chernyshov}}}, \bibinfo{journal}{Acta
  Cryst.} \textbf{\bibinfo{volume}{A64}}, \bibinfo{pages}{598}
  (\bibinfo{year}{2008}).

\bibitem[{\citenamefont{{T. Soma} and {H. Matsuo Kagaya}}(1983)}]{soma1983}
\bibinfo{author}{\bibnamefont{{T. Soma}}} \bibnamefont{and}
  \bibinfo{author}{\bibnamefont{{H. Matsuo Kagaya}}}, \bibinfo{journal}{phys.
  stat. sol. (b)} \textbf{\bibinfo{volume}{118}}, \bibinfo{pages}{245}
  (\bibinfo{year}{1983}).

\bibitem[{\citenamefont{{H.-M. Kagaya} and {T. Soma}}(1984)}]{kagaya1984}
\bibinfo{author}{\bibnamefont{{H.-M. Kagaya}}} \bibnamefont{and}
  \bibinfo{author}{\bibnamefont{{T. Soma}}}, \bibinfo{journal}{phys. stat. sol.
  (b)} \textbf{\bibinfo{volume}{121}}, \bibinfo{pages}{K113}
  (\bibinfo{year}{1984}).

\bibitem[{\citenamefont{{Shigeru Okamoto} and {Shinichi
  Sakurai}}(2003)}]{okamoto2003}
\bibinfo{author}{\bibnamefont{{Shigeru Okamoto}}} \bibnamefont{and}
  \bibinfo{author}{\bibnamefont{{Shinichi Sakurai}}}, \bibinfo{journal}{Journal
  of Applied Crystallography} \textbf{\bibinfo{volume}{36}},
  \bibinfo{pages}{982} (\bibinfo{year}{2003}).

\bibitem[{\citenamefont{{Eric R. Henry}}(1997)}]{henry1997}
\bibinfo{author}{\bibnamefont{{Eric R. Henry}}}, \bibinfo{journal}{Biophysics
  Journal} \textbf{\bibinfo{volume}{72}}, \bibinfo{pages}{652}
  (\bibinfo{year}{1997}).

\bibitem[{\citenamefont{{Marius Schmidt} et~al.}(2002)\citenamefont{{Marius
  Schmidt}, {Sudarshan Rajagopal}, {Zhong Ren}, and Moffat}}]{schmidt2002}
\bibinfo{author}{\bibnamefont{{Marius Schmidt}}},
  \bibinfo{author}{\bibnamefont{{Sudarshan Rajagopal}}},
  \bibinfo{author}{\bibnamefont{{Zhong Ren}}}, \bibnamefont{and}
  \bibinfo{author}{\bibfnamefont{K.}~\bibnamefont{Moffat}},
  \bibinfo{journal}{Biophysics Journal} \textbf{\bibinfo{volume}{84}},
  \bibinfo{pages}{2112} (\bibinfo{year}{2002}).

\bibitem[{\citenamefont{Debernardi}(1998)}]{debernardi1998}
\bibinfo{author}{\bibfnamefont{A.}~\bibnamefont{Debernardi}},
  \bibinfo{journal}{Phys. Rev. B} \textbf{\bibinfo{volume}{57}},
  \bibinfo{pages}{12847} (\bibinfo{year}{1998}).

\end{thebibliography}

\end{document}